\documentstyle[subeq,axodraw]{elsart}

\newcommand\vek[1]{\mbox{\rmfamily\bfseries\itshape#1}}
\newcommand\vekexp[1]{\mbox{\scriptsize\rmfamily\bfseries\itshape#1}}
\def\rd{{\rm d}}
\def\shft{\newline\phantom{}\quad}

\begin{document}
\begin{frontmatter}
\title{Another Derivation of a Sum Rule for the Two-Dimensional
       Two-Component Plasma}

\author{B. Jancovici},
\address{Laboratoire de Physique Th\' eorique~$^1$,
Universit\' e de Paris-Sud, B\^ atiment 210, \\
91405 Orsay Cedex, France;  E-mail: janco@th.u-psud.fr}
\thanks[CNRS]{Unit\' e Mixte de Recherche no. 8627-CNRS}

\thanks[VEGA]{Supported by Grant VEGA No. 2/4109/97}
\author{P. Kalinay\thanksref{VEGA}} and
\author{L. \v Samaj\thanksref{VEGA}}
\address{Institute of Physics, Slovak Academy of Sciences, D\' ubravsk\' a
cesta 9, \\ 842 28 Bratislava, Slovakia \\
 E-mail: (P.K.) fyzipali@nic.savba.sk,
(L.\v S) fyzimaes@nic.savba.sk}

\begin{abstract}
In a two-dimensional two-component plasma, the second moment of the
{\it number density} correlation function has the simple value
$\{ 12 \pi [1-(\Gamma/4)]^2\}^{-1}$, where $\Gamma$ is the dimensionless
coupling constant.
This result is derived directly by using diagrammatic methods.

\end{abstract}
\end{frontmatter}

{\it Dedicated to Joel Lebowitz on the occasion of his 70th birthday}

\section{Introduction}
\quad
The system under consideration is the two-dimensional ($2d$) two-component
plasma (TCP) (Coulomb gas), i.e. a system of positive and negative
point-particles of charge $\pm q$, in a plane. 
Two particles at a distance $r$ from each other interact through 
the $2d$ Coulomb interaction $\mp q^2 \ln (r/L)$, 
where $L$ is some irrelevant length (set to unity, for simplicity).
Classical equilibrium statistical mechanics is used. 
The dimensionless coupling constant is $\Gamma = \beta q^2$, where 
$\beta=1/kT$ is the inverse temperature. 
The system is known to be stable against collapse when $\Gamma < 2$. 
Let $\langle \hat n({\vek 0}) \hat n({\vek r})\rangle_T$ be the truncated 
density-density correlation function, where $\hat n({\vek r})$ is the total 
microscopic number density at ${\vek r}$.
\shft
In a recent paper \cite{1}, an exact expression for the second moment of
that density-density correlation function has been derived:
\begin{equation} \label{1}
\int \langle \hat n({\vek 0}) \hat n({\vek r}) \rangle_T ~ r^2~ \rd^2 {\vek r}
=  {1 \over 12 \pi \left( 1-{\Gamma\over 4} \right)^2} .
\end{equation}
However, the two alternative derivations which were given were rather
indirect ones, and the Conclusion of the paper had the sentence ``A~more
direct derivation is still wanted''. At about the same time, another
paper \cite{2} used a diagrammatic method for deriving an exact expression
for the sixth moment of the pair correlation function of the
$2d$ {\em one}-component plasma (OCP). The authors of both papers soon
realized that the diagrammatic method of Ref. 2 could be easily adapted
to the two-component case and used for providing a more direct
derivation of (\ref{1}), which is the main content of the present paper.
\shft
The paper is organized as follows. 
In Section 2, we generalize a renormalized Mayer expansion \cite{2}, \cite{3}, 
to the case of a many-component fluid. 
The application of the formalism to the $2d$ TCP in Section 3 leads, 
with the aid of the results obtained in Ref. \cite{2},
to the sum rule (\ref{1}) for the density correlation function.
The charge correlation function is discussed in Section 4.
Section 5 deals with consequences on finite-size effects for the sphere
geometry, confirming predictions from conformal invariance
\cite{4}, \cite{5}.

\section{Renormalized Mayer expansion for many-component fluids}
\quad
We first consider a general classical multi-component fluid
in thermodynamic equilibrium.
The last system is composed of distinct species of particles $\{ \sigma \}$
with the corresponding densities $\{ n({\vek r},\sigma) \}$ at
vector position ${\vek r}$ of a $d$-dimensio\-nal space.
The particles interact through the pair potential $v({\vek r}_i,\sigma_i
\vert {\vek r}_j,\sigma_j)$ which depends on the mutual distance
of particles $i, j$ as well as on their types $\sigma_i, \sigma_j$.
Vector ${\vek r}_i$ will be represented simply by $i$.
\shft
According to the Mayer diagrammatic formalism in density \cite{6},
the generalization from one-component to more-than-one component fluid 
consists in considering an additional particle-type label $\sigma$ 
for each, field (black) or root (white), vertex.
Simultaneously, besides the integration over spatial coordinate of
a field circle, the summation over all $\sigma$-states at this
vertex is assumed as well.
The renormalized Mayer representation of the excess Helmholtz free
energy \cite{2}, resulting
from the expansion of each Mayer function in the inverse temperature
and the consequent series elimination of two-coordinated field
circles between every couple of three- or more-coordinated field circles 
(by coordination of a circle we mean its bond-coordination,
i.e. the number of bonds meeting at this circle), can therefore be
straightforwardly extended to the many-component case.
The renormalized $K$-bonds are now given by
$$
{\begin{picture}(32,20)(0,7)
    \Photon(0,10)(32,10){1}{7}
    \BCirc(0,10){2.5} \BCirc(32,10){2.5}
    \Text(0,0)[]{$1,\sigma_1$} \Text(32,0)[]{$2,\sigma_2$}
    \Text(17,23)[]{$K$}
\end{picture}}
\ \ \ = \ \ \
{\begin{picture}(32,20)(0,7)
    \DashLine(0,10)(32,10){5}
    \BCirc(0,10){2.5} \BCirc(32,10){2.5}
    \Text(0,0)[]{$1,\sigma_1$} \Text(32,0)[]{$2,\sigma_2$}
    \Text(17,23)[]{$-\beta v$} 
\end{picture}}\ \ +\ \
{\begin{picture}(64,20)(0,7)
    \DashLine(0,10)(32,10){5}
    \DashLine(32,10)(64,10){5}
    \BCirc(0,10){2.5} \BCirc(64,10){2.5}
    \Vertex(32,10){2.2}
    \Text(0,0)[]{$1,\sigma_1$} \Text(64,0)[]{$2,\sigma_2$}
\end{picture}}\ \ +\ \
{\begin{picture}(96,20)(0,7)
    \DashLine(0,10)(32,10){5}
    \DashLine(32,10)(64,10){5}
    \DashLine(64,10)(96,10){5}
    \BCirc(0,10){2.5} \BCirc(96,10){2.5}
    \Vertex(32,10){2.2} \Vertex(64,10){2.2}
    \Text(0,0)[]{$1,\sigma_1$} \Text(96,0)[]{$2,\sigma_2$}
\end{picture}}\ \ + \ldots$$

or, algebraically,
\begin{eqnarray} \label{2}
K(1,\sigma_1\vert 2,\sigma_2)& = & [-\beta v(1,\sigma_1 \vert 2, \sigma_2)]
\nonumber \\
& & + \sum_{\sigma_3} \int [-\beta v(1,\sigma_1 \vert 3,\sigma_3)]~
n(3,\sigma_3)~ K(3,\sigma_3 \vert 2,\sigma_2)~ \rd 3 .
\end{eqnarray}
The procedure of the bond-renormalization transforms the ordinary graphical
Mayer representation of the free energy into
\begin{subequations} \label{3}
\begin{equation} \label{3a}
\beta \bar F^{ex}[n] = \ \  
\begin{picture}(50,20)(0,7)
    \DashLine(0,10)(40,10){5}
    \Vertex(0,10){2.2} \Vertex(40,10){2.2}
\end{picture}
 + \ D^{(0)}[n] + \ \sum_{s=1}^{\infty}\ D^{(s)}[n] ,
\end{equation}
where ${\bar F}^{ex}$ is minus the excess free energy,
\begin{equation} \label{3b}
D^{(0)} = \ \
\begin{picture}(40,20)(0,7)
    \DashCArc(20,-10)(28,45,135){5}
    \DashCArc(20,30)(28,225,315){5}
    \Vertex(0,10){2} \Vertex(40,10){2}
\end{picture} \ \ +\ \ 
\begin{picture}(40,20)(0,19)
    \DashLine(0,10)(40,10){5}
    \DashLine(0,10)(20,37){5}
    \DashLine(20,37)(40,10){5}
    \Vertex(0,10){2} \Vertex(40,10){2} \Vertex(20,37){2}
\end{picture} \ \ +\ \ 
\begin{picture}(30,20)(0,10)
    \DashLine(0,0)(30,0){5}
    \DashLine(0,0)(0,30){5}
    \DashLine(0,30)(30,30){5}
    \DashLine(30,0)(30,30){5}
    \Vertex(0,0){2} \Vertex(30,0){2} \Vertex(0,30){2} \Vertex(30,30){2}
\end{picture} \ \ +\ \ \ldots 
\end{equation}
is the sum of all unrenormalized ring diagrams (which cannot undertake
the renormalization procedure because of the absence of three- or
more-coordinated field points) and
\begin{eqnarray} \label{3c} 
\sum_{s=1}^{\infty} D^{(s)} & = & \big\{
{\rm all\ connected\ diagrams\ which\ consist\ of\ } N\ge 2 {\rm
\ field} \nonumber \\
& & n(i,\sigma_i){\rm -circles\ of\ (bond)\ coordination\ }\ge 3 
{\rm \ and\ multiple }  \nonumber \\
& &K(i,\sigma_i\vert j,\sigma_j){\rm -bonds,\ and\ are\ free\ 
of\ connecting\ circles\  \big\}}
\end{eqnarray}
\end{subequations}
represents the set of all remaining completely renormalized graphs.
By multiple $K$-bonds we mean the possibility of an arbitrary
number of $K$-bonds between a couple of field circles, with the
topological factor $1/({\rm number\ of\ bonds})! $. 
The order of $s$-enumeration is irrelevant, let us say
\begin{equation} \label{4}
\begin{picture}(60,40)(0,7)
    \PhotonArc(20,6)(20,15,165){1}{11}
    \PhotonArc(20,14)(20,195,345){1}{11}
    \Photon(0,10)(40,10){1}{8.5}
    \Vertex(0,10){2} \Vertex(40,10){2}
    \Text(20,-25)[]{$D^{(1)}$}
\end{picture}
\begin{picture}(60,40)(0,7)
    \PhotonArc(20,-10)(28,45,135){1}{9}
    \PhotonArc(20,30)(28,225,315){1}{9}
    \PhotonArc(20,6)(20,15,165){1}{11}
    \PhotonArc(20,14)(20,195,345){1}{11}
    \Vertex(0,10){2} \Vertex(40,10){2}
    \Text(20,-25)[]{$D^{(2)}$}
\end{picture}
\begin{picture}(60,40)(0,16)
    \PhotonArc(32,6)(32,115,170){1}{7.5}
    \PhotonArc(-12,40)(32,295,355){1}{7.5}
    \PhotonArc(9,5)(32,10,70){1}{7.5}
    \PhotonArc(52,40)(32,185,250){1}{7.5}
    \Photon(0,10)(40,10){1}{8}
    \Vertex(0,10){2} \Vertex(40,10){2} \Vertex(20,35){2}
    \Text(20,-15)[]{$D^{(3)}$}
\end{picture}
\SetScale{0.9}
\begin{picture}(60,40)(0,13)
    \Photon(0,0)(0,40){1}{7}
    \Photon(40,0)(40,40){1}{7}
    \Vertex(0,0){2} \Vertex(40,0){2} \Vertex(0,40){2} \Vertex(40,40){2}
    \PhotonArc(20,-20)(28,45,135){1}{8}
    \PhotonArc(20,20)(28,225,315){1}{8}
    \PhotonArc(20,20)(28,45,135){1}{8}
    \PhotonArc(20,60)(28,225,315){1}{8}
    \Text(20,-19)[]{$D^{(4)}$}
\end{picture} 
\begin{picture}(60,40)(0,12)
    \Photon(0,0)(40,0){1}{7}
    \Photon(0,0)(0,40){1}{7}
    \Photon(0,40)(40,40){1}{7}
    \Photon(40,0)(40,40){1}{7}
    \Photon(0,0)(40,40){1}{8}
    \Photon(0,40)(40,0){1}{8}
    \Vertex(0,0){2} \Vertex(40,0){2} \Vertex(0,40){2} \Vertex(40,40){2}
    \Text(20,-19)[]{$D^{(5)}$}
\end{picture}
\SetScale{1}
\end{equation}
\vskip1.5cm
etc.
\shft
The truncated pair correlation 
\begin{equation} \label{5}
h(1,\sigma_1\vert 2,\sigma_2) = {n_2(1,\sigma_1\vert 2,\sigma_2)
\over n(1,\sigma_1) n(2,\sigma_2)} -1
\end{equation}
is related to the pair direct correlation $c$ via 
the Ornstein-Zernike (OZ) equation
\begin{eqnarray} \label{6}
h(1,\sigma_1\vert 2,\sigma_2) & = & c(1,\sigma_1\vert 2,\sigma_2)
\nonumber \\
& & + \sum_{\sigma_3} \int c(1,\sigma_1\vert 3,\sigma_3)
n(3,\sigma_3) h(3,\sigma_3\vert 2,\sigma_2) \rd 3 .
\end{eqnarray}
The free energy is the generating functional for $c$ in the sense that
\begin{equation} \label{7}
c(1,\sigma_1\vert 2,\sigma_2) = {\delta^2 (\beta \bar F^{ex})
\over \delta n(1,\sigma_1) \delta n(2,\sigma_2)} .
\end{equation}
According to (\ref{3}), the direct correlation is thus written as
\begin{subequations} \label{8}
\begin{equation} \label{8a}
c(1,\sigma_1\vert 2,\sigma_2) = \ \ 
\begin{picture}(35,20)(0,7)
    \DashLine(0,10)(35,10){5}
    \BCirc(0,10){2.5} \BCirc(35,10){2.5}
    \Text(0,0)[]{$1,\sigma_1$} \Text(35,0)[]{$2,\sigma_2$}
\end{picture}\ \ +
c^{(0)}(1,\sigma_1\vert 2,\sigma_2) + \sum_{s=1}^{\infty} 
c^{(s)}(1,\sigma_1\vert 2,\sigma_2) ,
\end{equation}
where $c^{(0)}(1,\sigma_1\vert 2,\sigma_2) = \delta^2 D^{(0)} / 
\delta n(1,\sigma_1) \delta n(2,\sigma_2)$ can be readily shown 
to correspond to the renormalized Meeron graph
\begin{equation} \label{8b}
c^{(0)}(1,\sigma_1\vert 2,\sigma_2) = \ \ 
\begin{picture}(50,25)(0,7)
    \PhotonArc(20,-10)(28,45,135){1}{9}
    \PhotonArc(20,30)(28,225,315){1}{9}
    \BCirc(0,10){2.5} \BCirc(40,10){2.5}
    \Text(0,0)[]{$1,\sigma_1$} \Text(44,0)[]{$2,\sigma_2$}
\end{picture}\ \ = {1\over 2!} K^2(1,\sigma_1\vert 2,\sigma_2)
\end{equation}
and 
\begin{equation} \label{8c}
c^{(s)}(1,\sigma_1\vert 2,\sigma_2) = \displaystyle{
\delta^2 D^{(s)} \over \delta n(1,\sigma_1) \delta n(2,\sigma_2)} 
\end{equation}
\end{subequations}
$(s=1,2,\ldots)$ denotes the whole family of 
$(1,\sigma_1)(2,\sigma_2)$-rooted diagrams generated from $D^{(s)}$.
To get explicitly a direct-correlation family, one has to take into
account the functional dependence of the dressed $K$-bonds (\ref{2}) 
on the density, too. 
It is straightforward to verify by variation of (\ref{2}) that it holds
\begin{equation} \label{9}
{\delta K(1,\sigma_1\vert 2,\sigma_2) \over \delta n(3,\sigma_3)} = 
K(1,\sigma_1\vert 3,\sigma_3) K(3,\sigma_3 \vert 2,\sigma_2) . 
\end{equation}
As a consequence, the functional derivative of $D^{(s)}$ with respect to the 
density generates root-circles, besides the field-circle positions,
also on $K$-bonds, causing their ``correct'' K-K division.
For example, in the case of the generator $D^{(1)}$ drawn in (\ref{4}),
we obtain
\begin{eqnarray} \label{10} 
c^{(1)}(1,\sigma_1\vert 2,\sigma_2)=\ \ \ \ \
\begin{picture}(55,40)(0,7)
    \PhotonArc(20,6)(20,15,165){1}{11}
    \PhotonArc(20,14)(20,195,345){1}{11}
    \Photon(0,10)(40,10){1}{8.5}
    \BCirc(0,10){2.5} \BCirc(40,10){2.5}
    \Text(-7,-2)[]{$1,\sigma_1$} \Text(49,-2)[]{$2,\sigma_2$}
\end{picture} +\ \ \ \ \
\begin{picture}(55,40)(0,7)
    \PhotonArc(20,6)(20,15,165){1}{11}
    \PhotonArc(20,14)(20,195,345){1}{11}
    \Photon(0,10)(40,10){1}{8.5}
    \BCirc(0,10){2.5} \BCirc(20,26){2.5}
    \Text(-7,-2)[]{$1,\sigma_1$} \Text(20,36)[]{$2,\sigma_2$} 
    \Vertex(40,10){2.2}
\end{picture} +\ \ \ \ \
\begin{picture}(55,40)(0,7)
    \PhotonArc(20,6)(20,15,165){1}{11}
    \PhotonArc(20,14)(20,195,345){1}{11}
    \Photon(0,10)(40,10){1}{8.5}
    \BCirc(0,10){2.5} \BCirc(20,26){2.5}
    \Text(-7,-2)[]{$2,\sigma_2$} \Text(20,36)[]{$1,\sigma_1$} 
    \Vertex(40,10){2.2}
\end{picture} \nonumber \\
+\ \ \
\begin{picture}(55,50)(0,7)
    \PhotonArc(20,6)(20,15,165){1}{11}
    \PhotonArc(20,14)(20,195,345){1}{11}
    \Photon(0,10)(40,10){1}{8.5}
    \BCirc(20,26){2.5} \BCirc(20,-6){2.5}
    \Vertex(0,10){2.2} \Vertex(40,10){2.2}
    \Text(20,36)[]{$1,\sigma_1$} \Text(20,-16)[]{$2,\sigma_2$}
\end{picture}\ +\ \ \
\begin{picture}(55,50)(0,7)
    \PhotonArc(20,6)(20,15,165){1}{11}
    \PhotonArc(20,14)(20,195,345){1}{11}
    \Photon(0,10)(40,10){1}{8.5}
    \BCirc(11,24){2.5} \BCirc(29,24){2.5}
    \Vertex(0,10){2.2} \Vertex(40,10){2.2}
    \Text(6,34)[]{$1,\sigma_1$} \Text(34,34)[]{$2,\sigma_2$}
\end{picture} 
\end{eqnarray}

\section{Density correlations in the 2d TCP}
\quad
Let us now concentrate on the neutral $2d$ TCP of positive $(\sigma = +)$
and negative $(\sigma = -)$ charges, with the Coulomb interaction
energy given by
\begin{equation} \label{11}
-\beta v(i,\sigma_i\vert j,\sigma_j) = \sigma_i \sigma_j \Gamma
\ln \vert i-j \vert .
\end{equation}
We consider the infinite-volume limit, characterized by homogeneous
densities $n(1,\sigma) = n_{\sigma}$ and both isotropic and
translationally invariant two-body correlations,
$c(1,\sigma_1\vert 2,\sigma_2) = c_{\sigma_1 \sigma_2}(\vert 1-2 \vert),
h(1,\sigma_1\vert 2,\sigma_2) = h_{\sigma_1 \sigma_2}(\vert 1-2 \vert).$
The requirement of the charge neutrality, $n_+ = n_- = n/2$ ($n$ is
the total number density of particles), automatically implies the symmetry 
with respect to the $+\leftrightarrow -$ interchange of the charge,
\begin{subequations} \label{12}
\begin{equation} \label{12a}
h_{++} = h_{--} \ ( = h_+ ), \quad \quad h_{+-} = h_{-+} \ ( = h_- ) ;
\end{equation}
\begin{equation} \label{12b}
c_{++} = c_{--} \ ( = c_+ ), \quad \quad c_{+-} = c_{-+} \ ( = c_- ) .
\end{equation}
\end{subequations}
Thus, in Fourier space
\begin{subequations} \label{13}
\begin{equation} \label{13a}
f({\vek r}) = {1\over 2\pi} \int \exp(i{\vek k}\cdot 
{\vek r}) \hat f({\vek k}) \rd^2 {\vek k} ,
\end{equation}
\begin{eqnarray} \label{13b}
\hat f({\vek k}) & = & {1\over 2\pi} \int \exp(-i{\vek k}
\cdot {\vek r}) f({\vek r}) \rd^2 {\vek r} \nonumber \\
& = & \sum_{s=0}^{\infty} {(-1)^s\over (s!)^2} 
\left( {k^2\over 4} \right)^s 
{1\over 2\pi} \int f(r) r^{2s} \rd^2 {\vek r} ,
\end{eqnarray}
\end{subequations}
the OZ equation (\ref{6}) implies
\begin{subequations} \label{14}
\begin{eqnarray}
\hat h_+(k) & = & \hat c_+(k)\left[ 1 + \pi n \hat h_+(k) \right] + \pi n
\hat c_-(k) \hat h_-(k) , \label{14a} \\
\hat h_-(k) & = & \hat c_-(k)\left[ 1 + \pi n \hat h_+(k) \right] + \pi n
\hat c_+(k) \hat h_-(k) , \label{14b}
\end{eqnarray}
\end{subequations}
where $k = \vert {\vek k} \vert$.
Introducing
\begin{subequations} \label{15}
\begin{eqnarray}
h & = & {1\over 4} \sum_{\sigma, \sigma' = \pm} h_{\sigma \sigma'} = 
{1\over 2} ( h_+ + h_-) , \label{15a} \\
c & = & {1\over 4} \sum_{\sigma, \sigma' = \pm} c_{\sigma \sigma'} = 
{1\over 2} ( c_+ + c_-) , \label{15b}
\end{eqnarray}
\end{subequations}
the summation of (\ref{14a}) and (\ref{14b}) yields
\begin{equation} \label{16}
\hat h(k) = \hat c(k) + 2 \pi n \hat c(k) \hat h(k) .
\end{equation}
\quad
In the renormalized format, due to the translational invariance of
renormalized bonds $K(1,\sigma_1\vert 2,\sigma_2) = K_{\sigma_1 \sigma_2}
(\vert 1-2 \vert)$, the Fourier transformation of relation (\ref{2}) results in
\begin{equation} \label{17}
\hat K_{\sigma \sigma'}(k) = [- \beta \hat v_{\sigma \sigma'}(k)]
+ 2\pi \sum_{\sigma''=\pm} \left[ -\beta \hat v_{\sigma \sigma''}(k)\right]
n_{\sigma''} \hat K_{\sigma'' \sigma'}(k) .
\end{equation}
With regard to
\begin{equation} \label{18}
-\beta \hat v_{\sigma \sigma'}(k) = -\sigma \sigma' \Gamma /k^2 ,
\end{equation}
equation (\ref{17}) yields
\begin{subequations} \label{19}
\begin{eqnarray}
\hat K_{\sigma \sigma'}(k) & = & - \sigma \sigma'
{\Gamma \over k^2 + 2\pi \Gamma (n_+ + n_-)} , \label{19a}\\
K_{\sigma \sigma'}(r) & = & - \sigma \sigma'
\Gamma K_0\left[ r\sqrt{2\pi\Gamma(n_+ + n_-)}\right] , \label{19b}
\end{eqnarray}
\end{subequations}
with $K_0$ being the modified Bessel function of second kind.
The pair direct correlation (\ref{8}) is expressible as
\begin{eqnarray} \label{20}
c_{\sigma_1 \sigma_2}(\vert 1-2 \vert) & = &
\sigma_1 \sigma_2 \ln \vert 1-2 \vert
+ {1\over 2} \left[ \Gamma K_0(\vert 1-2 \vert \sqrt{2\pi \Gamma n})
\right]^2 \nonumber \\  
& & + \sum_{s=1}^{\infty} 
c^{(s)}_{\sigma_1 \sigma_2}(\vert 1-2 \vert) .
\end{eqnarray}
$c$, equation (\ref{15b}), is then given by
\begin{subequations} \label{21}
\begin{equation} \label{21a}
c(\vert 1-2 \vert) = {1\over 2} \left[ \Gamma K_0(\vert 1-2 \vert 
\sqrt{2\pi \Gamma n}) \right]^2 + \sum_{s=1}^{\infty} 
c^{(s)}(\vert 1-2 \vert) ,
\end{equation}
where
\begin{eqnarray} \label{21b}
c^{(s)}(\vert 1-2 \vert) & = & {1\over 4} \sum_{\sigma_1,\sigma_2 = \pm}
c^{(s)}_{\sigma_1 \sigma_2}(\vert 1-2 \vert) \nonumber \\
& = & {1\over 4}
\sum_{\sigma_1,\sigma_2 = \pm} {\delta^2 D^{(s)} \over \delta n(1,\sigma_1)
\delta n(2,\sigma_2)}\Big\vert_{n({\vekexp r},\sigma) = n/2} .
\end{eqnarray}
\end{subequations}
\shft
As concerns the evaluation of $c^{(s)}$, because of the symmetry
$K_{\sigma, \sigma} = - K_{\sigma,-\sigma}$, it is rather obvious
that for $n_+=n_-$ the summation on signs gives a non-vanishing $c^{(s)}$
only if all the vertices of the generator $D^{(s)}$ have an even
bond-coordination [note that after the functional derivation (21b),
with regard to (9),
all the white circles will also have an even bond-coordination].
Furthermore the resulting $c^{(s)}$ is the same as in the case
of the OCP with particle density $n_+ + n_- = n$.
It was proven in Ref. \cite{2} that the zeroth- and second-moments of
such a $c^{(s)}$ vanish independently of the topology of $D^{(s)}$,
so that
\begin{equation} \label{22}
\hat c^{(s)}(k) = O(k^4) .
\end{equation}
\quad
To summarize, using formula \cite{7}
\begin{equation} \label{23}
\int_0^{\infty} K_0^2(x) x^{1+2s} \rd x = 2^{(2s-1)} {(s!)^4 \over
(2s+1)!} \quad \quad (s\ge 0)
\end{equation}
in the Fourier representation (\ref{13b}) of $K_0^2$ and regarding equation 
(\ref{22}), the small-$k$ expansion
of the Fourier component of $c$ (\ref{21}) reads
\begin{equation} \label{24}
\hat c(k) = {\Gamma \over 8 \pi n} - {k^2 \over 96 (\pi n)^2}
+ O(k^4) .
\end{equation}
Inserting (\ref{24}) into (\ref{16}), we find
\begin{equation} \label{25}
\hat h(k) = {\Gamma \over 8 \pi n \left( 1-{\Gamma \over 4} \right)}
- {k^2 \over 96 (\pi n)^2 \left( 1-{\Gamma \over 4} \right)^2} + O(k^4) .
\end{equation}
The truncated density-density correlation
 
is related to $h$ [see definitions (\ref{5}), (\ref{12a}) and (\ref{15a})] via
\begin{equation} \label{26}
\langle \hat n({\vek 0}) \hat n({\vek r}) \rangle_T =
n^2 h(r) + n \delta({\vek r}) .
\end{equation}
With respect to (\ref{13b}), we finally obtain
\begin{subequations} \label{27}
\begin{eqnarray}
n \int h({\vek r})\rd^2 {\vek r}
& = & {\Gamma \over 4\left( 1-{\Gamma\over 4} \right)} , \label{27a} \\
\int \langle \hat n({\vek 0}) \hat n({\vek r}) \rangle_T ~ r^2~ \rd^2 {\vek r}
& = & {1 \over 12 \pi \left( 1-{\Gamma\over 4} \right)^2} . \label{27b}
\end{eqnarray}
\end{subequations}
The first formula (\ref{27a}) is identical to the compressibility sum rule
\begin{equation} \label{28} 
n \int h({\vek r}) 
\rd^2 {\vek r} = {\partial n \over \partial (\beta p)} - 1
\end{equation}
with the use of the exact equation of state
$\beta p = n ( 1- \Gamma /4)$, where $p$ is the pressure.
The second formula (\ref{27b}) coincides with the new sum rule (\ref{1}) 
suggested in Ref. \cite{1}.

\section{Charge correlations in the 2d TCP}
\quad
One might try to use the present formalism for studying the charge
correlations. 
Unfortunately, one can only reproduce known results. 
Indeed, introducing the charge correlation and direct
correlation functions         
\begin{equation} \label{29}
h' =  {1\over 2} ( h_+ - h_-) , \quad \quad
c' =  {1\over 2} ( c_+ - c_-) ,
\end{equation}
one now finds that the non-vanishing graphs contributing to $c'$ are the
ones which have their two root-vertices with an odd coordination and
their field vertices still with an even coordination. 
For instance, only the first diagram in the $c^{(1)}$ family [see formula
(10)] survives.
The argument of Ref. \cite{2} no longer applies, and the corresponding 
contribution to $c'(k)$ is $O(1)$ rather than $O(k^4)$. 
One finds     
\begin{equation} \label{30}
c'(k) = - \Gamma/k^2 + O(1)
\end{equation}
and this leads only to the known Stillinger-Lovett sum rules for the
zeroth and second moments of the charge correlation function $h'$.

\section{Consequences}
\quad
Cranking backwards the arguments in Ref. \cite{1}, it is now possible to
derive conformal-invariance properties from the 2nd moment (\ref{27b}). 
Indeed, within the scheme established in Ref. \cite{1}, 
the logarithm of the grand-canonical partition function of the 2d TCP, 
formulated on the surface of a $3d$ sphere of radius $R$, can be written as
\begin{subequations} \label{31}
\begin{equation} \label{31a}
\ln \Xi = 4 \pi R^2 \beta p + f(R) ,
\end{equation}
where, in the limit $R\to \infty$,
\begin{equation} \label{31b}
{\partial f(R) \over \partial R} = - {1\over R} \left( 1-
{\Gamma \over 4} \right)^2 4\pi \int 
\langle \hat n({\vek 0}) \hat n({\vek r})\rangle_T ~ r^2~ \rd^2 {\vek r} .
\end{equation}
\end{subequations}
Assuming the sum rule (\ref{27b}), one finds the behaviour, as $R\to\infty$,
\begin{equation} \label{32}
f(R) \sim - {1\over 3} \ln R + {\rm const} ,
\end{equation}
i.e., for an arbitrary coupling $\Gamma$, the finite-size correction 
to $\ln \Xi$ exhibits the universal form suggested by using
predictions of the conformal-invariance theory \cite{4}, \cite{5},
with the correct prefactor for the sphere geometry.
\shft
The mapping of the $2d$ TCP onto the massive Thirring model confirms
the validity of a sum rule for the latter model 
(for details, see Section 4 of Ref. \cite{1}).

\section{Conclusion}
\quad
Unlike previously known moments in Coulomb systems, the new 6th moment
of the pair correlation function for the one-component plasma, and the
new 2nd moment of the density (not charge) correlation function for the
two-component plasma have been exactly obtained only for two-dimensions
and point-charges (no short-range interactions). A~simple basic ingredient
of the derivations is that these systems have only one length scale 
$n^{-1/2}$ provided by the density; the quantities of interest depend on the
density only trivially through that length scale. Nevertheless,
technically, showing that higher order graphs give to $c$ contributions
$O(k^4)$ (Ref. \cite{2}) is somewhat lengthy. Whether a simpler
derivation is possible is an open problem.

\end{document}